\documentclass{article}

\PassOptionsToPackage{numbers, compress}{natbib}

\usepackage[final]{neurips_2021}

\usepackage[utf8]{inputenc} 
\usepackage[T1]{fontenc}    
\usepackage{hyperref}       
\usepackage{url}            
\usepackage{booktabs}       
\usepackage{amsfonts}       
\usepackage{nicefrac}       
\usepackage{microtype}      
\usepackage{xcolor}         
\usepackage{graphicx}
\usepackage{wrapfig}

\title{Malakai: Music That Adapts to the Shape of Emotions}

\author{%
  \textbf{Zack Harris, Liam Atticus Clarke, Pietro Gagliano,} \\ 
  \textbf{Dante Camarena, Manal Siddiqui} \\
  Transitional Forms \\
  \url{https://transforms.ai/}\\
  \texttt{technology@transforms.ai} \\
   \And
  \textbf{Pablo S. Castro} \\ 
     \texttt{Google Brain} \\
   \texttt{psc@google.com} \\
}

\begin{document}

\maketitle

\section{Background}

This is a strange and exciting time for computer-generated music. The idea of computer-generated musical composition has captured the public imagination, as far back as Kurzweil’s demonstration of a pattern-based composer on live TV in 1965\cite{academyofachievement_2019}. Since then, improvements in technology and composition tools have created whole musical genres based around computer-generated compositions, and have resulted in a vast library of algorithmic compositional techniques. Furthermore, in the past few decades, interactive media such as games and virtual reality have resulted in a demand for music that can adapt to dynamic circumstances presented within the interactive medium.  Finally, the advent of ML music models such as Google Magenta’s MusicVAE\cite{conf/icml/RobertsERHE18} now allow us to extract and replicate compositional features from otherwise complex datasets. These models allow computational composers to parameterize abstract variables such as style and mood.

By leveraging these models and combining them with procedural algorithms from the last few decades, it is possible to create a \textbf{dynamic song}  that composes music in real-time to accompany interactive experiences \cite{zulic2019ai}. \textbf{Malakai} is a tool that helps users of varying skill levels create, listen to, remix and share such dynamic songs. Using Malakai, a Composer can create a dynamic song that can be interacted with by a Listener.

\section{Project Malakai}

Malakai is comprised of three main components, The Curves, The Graph and The Blocks:

\paragraph{The Curves}
A dynamic music player (\autoref{fig:curves}). As opposed to conventional music tracks, Malakai tracks are dynamically composed based on certain input parameters. We have developed a simple player interface that allows listeners to define the value of these parameters over the course of the song by modifying a curve. This interface takes inspiration from storytelling curves.\cite{vonnegut_1987, vonnegut_2010} The intended user of this interface is the Listener.
By default, tracks that we developed internally are based on emotional values. These values correspond with the three-variable model introduced by \cite{greenberg}. Listeners are able to modulate Energy, Valence and Complexity (Matching Greenberg’s Arousal, Valence and Depth). For example, higher valence produces happier or brighter sounding chord progressions, while lower valence leads to progressions with a darker mood.

\begin{wrapfigure}[16]{r}{0.3\textwidth}
  \begin{center}
    \includegraphics[width=0.3\textwidth]{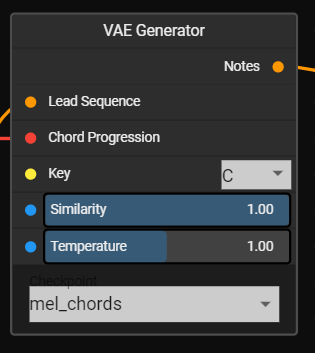}
    \caption{A MusicVAE Block}
    \label{fig:block}
  \end{center}
\end{wrapfigure}

\paragraph{The Blocks} 
A Block (\autoref{fig:block}) is a musical process, which represents a transformation or generation of musical elements. Each Block has a set of inputs and outputs. One example would be a block that improvises a melody according to an input chord progression that was generated by another block. To ensure that the user cannot make systematic mistakes (plugging incompatible blocks together), blocks show coloured ports to represent the type of date expected. E.g. Symbolic Music (Notes, rhythm, progressions, etc), Audio data (raw sound, filters, etc.) or others. Blocks can vary between simple transformations to implementations of Machine Learning models.

In order to effectively create dynamic songs, the Composer must become familiar with the output of the different ML models available through Malakai. Each AI model has a block that allows configuration of its options. We found that this kind of low level control, combined with real-time interaction has allowed our musicians to create music that accentuates each model’s peculiarities. Furthermore, we created a series of procedural emotional music generators for different musical elements, such as chord progressions, rhythms and melodies. These generators can have their emotional parameters set manually, or can receive the emotional curve values as inputs which change over the duration of the song.

\paragraph{The Graph} 
An authoring tool for dynamic songs.(\autoref{fig:graph}) This tool allows a user to combine Machine Learning and algorithmic music generators to author the dynamic songs that can be heard in the player.  This goal is achieved through a node graph interface inspired by other no-code programming tools such as Node-Based Shader editors \cite{gamegarage, shadergraph}. The intended user of this interface is the Composer. 

Using the graph, Composers can drag, drop, and connect blocks together to build a dynamic song. In doing so, the Composer can provide their artistic intent by selecting the instrumentation, progressions, melodies and structure. Composers may provide these statically (through an instrument or midi file) or by using the Algorithmic or ML generators. 

By using this system, the Composer is encouraged to construct an abstraction of their music in such a way that can be interacted with by the listener through the Curve interface. In a way, this form of composition is uniquely distinct from traditional composition, as it requires collaboration with both the user and the AI models (in which performance varies due to stochasticity). As a result, the composer is invited to merely set the stage for the performance, allowing the user and AI models to participate in the composition, a process that we label Three-Way Authorship \cite{agence}. 

\section{Example}

\begin{figure}[h!]
\centering
  \centering
  \includegraphics[width=\linewidth]{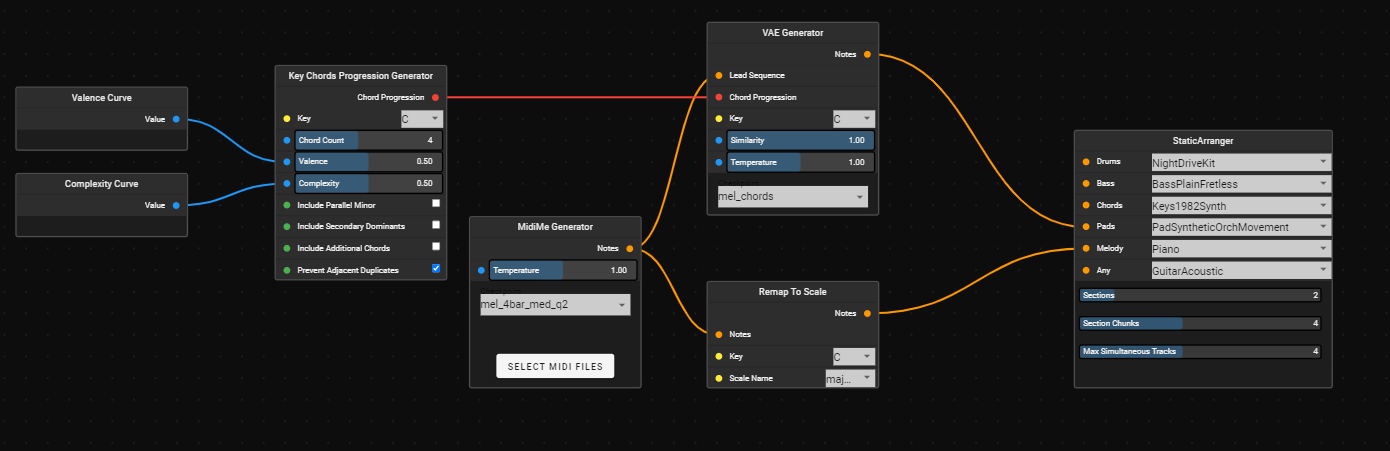}
  \caption{A simple dynamic song}
  \label{fig:graph}
\end{figure}

In our example below (\autoref{fig:graph}), the melody was provided with MidiMe \cite{midime}, which trained a MusicVAE model within a given latent space. This provided the leading melody of the performance. That output was then also mapped to a separate MusicVAE that provided the countermelody. That countermelody made use of MusicVAE’s underlying chords feature to match the chords that were created by an algorithmic progression generator. That generator was in turn fed by the user input of the curve parameters. This performance was then able to vary in tension by modulating the counter-melody and chord progressions by using the valence and complexity parameters. We’ve also created a preliminary system which sets the foundation for allowing dynamic tracks to be embedded in games, with the parameters tied to the state of the game. This will result in an atmospheric track that mirrors the user’s emotional state within the context of the game. This kind of interactive composition will be more important as interactive media continues to increase in popularity.

\pagebreak
\appendix

\section{Appendix}

\subsection{Ethical Considerations}

There are foundational ethical questions related to machine learning and creativity in music. For example, it is not clear if machine-generated works could, should and will be protected by Intellectual Property, and, if they do, who should be the owner of the related rights. In addition, other problems related to copyright should be considered in our opinion, such as if training over protected work is admitted and what happens if the results are similar to some of the training data. Another ongoing important debate is about authorship and the human role in creative fields in the era of AI.

\subsection{Technical description}

Malakai is powered by Tensorflow.js and renders audio using the WebAudio API. It can also be used as a standalone composition library. Using the Malakai API, it can be extended to support new blocks. While the focus on the application is the use of AI blocks, Procedural blocks allow for more controllable transformations of data and the opportunity for creative coders to create new musical algorithmic generators as new blocks in our modular system. As such, we leverage the existing high level audio library Tone.js and music theory library tonal.js, which provide an easy to use way for users to write compositional elements.

Malakai is currently in a functional state, with a few remaining bugs. If selected, Malakai will be made available to NeurIPS audience as a browser-run application. If chosen for a presentation, the presentation will musical samples.

\subsection{Acknowledgements}

Thanks to Kory Mathewson, Liam Clarke (unrelated to author), Alexander Bakogeorge, Anna Huang, Erin Ray, and the rest of the Transforms.ai team for their help and/or advice in conducting this research.

\subsection{The curves interface}

\begin{figure}[h!]
\centering
  \centering
  \includegraphics[width=\linewidth]{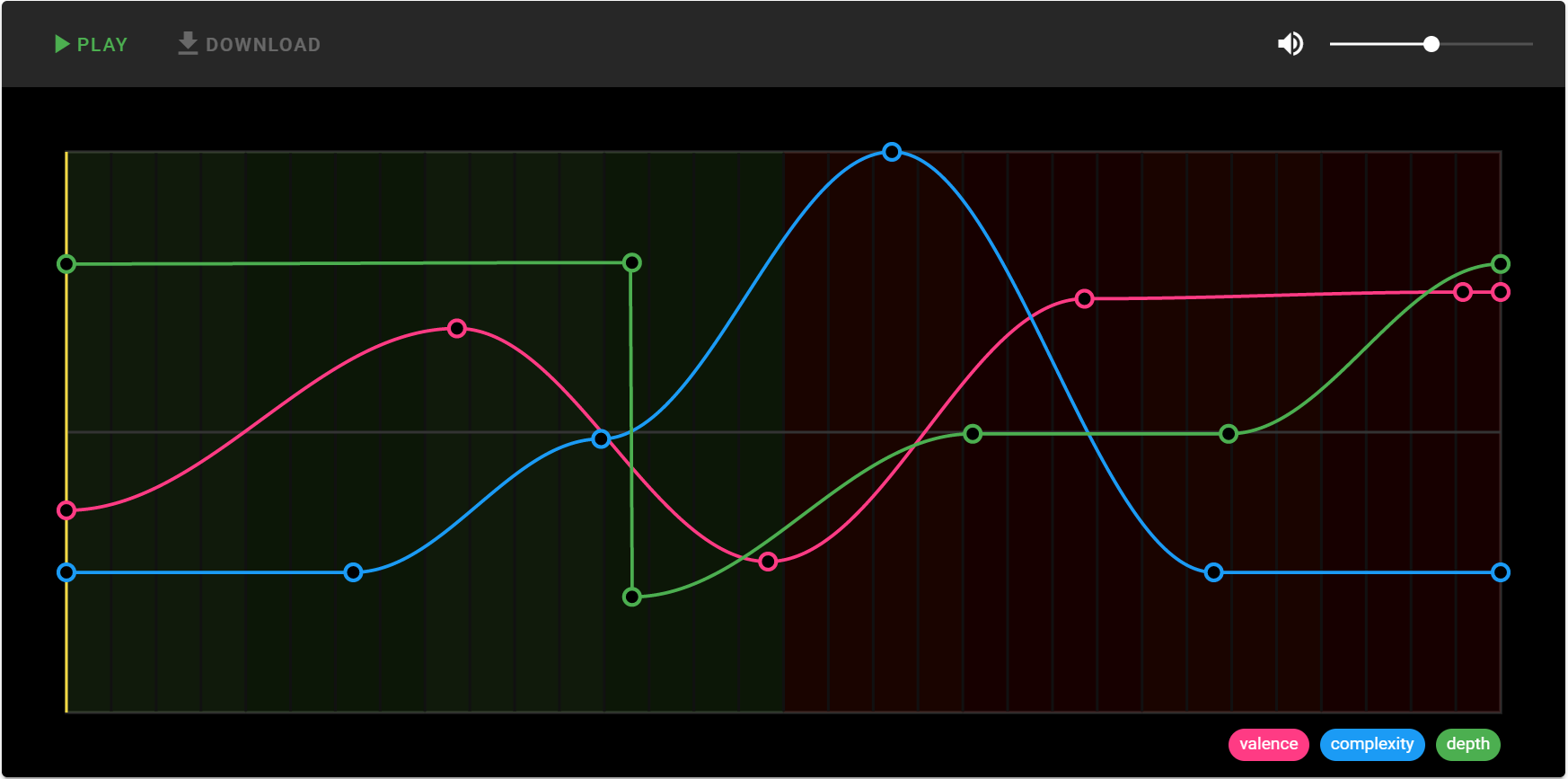}
  \caption{The curves interface}
  \label{fig:curves}
\end{figure}

\pagebreak
\bibliographystyle{abbrvnat}
\bibliography{references}

\end{document}